\documentstyle{article}

\author{Zhen Wang\\
Physics Department, LiaoNing Normal University, DaLian 116029, P. R. China}
\title{Calculation of the Deflection of Light Ray near the Sun with Quantum-corrected Newton's Gravitation Law
}
\input tcilatex

\QQQ{Subject}{
}

\QQQ{Keywords}{
deflection of light ray, quantum
}

\QQQ{Language}{
American English
}

\begin{document}

\maketitle
\begin{abstract}
The deflection of light ray passing near the Sun is calculated with
quantum-corrected Newton's gravitation law. The satisfactory result suggests
that there may exist other theoretical possibilities besides the theory of
relativity.
\end{abstract}

The deflection of light ray near the Sun was a successful prediction of
Einstein's theory of relativity, to which classical mechanics is futile.
That was one of the best experimental supports to Einstein's theory. But as
many people know, the successful synthesis of Einstein's theory with quantum
theory has not emerged. In the attempt to construct the theory of quantum
gravity, some scientists even thought that one of the two theories was
temporary[1]. Though research on such a theory of quantum gravity has made
no convincing breakthrough, some features of it can be seen[2]. One of the
crucial points is the quantization of spacetime.

To get a theory which is compatible with quantum mechanics and at the same
time preserves the successful conclusions of the theory of relativity as
much as possible, I proposed a theoretical framework in recent years[3-5].
It has given new insights to problems like EPR paradox. Also, consideration
of space quantization in this framework gives naturally a correction to
Newton's Gravitation Law[6]:

$$
F=G\frac{Mm}{r(r-\delta )}\;\;\;\;\;\;\;\;\;\;\;\;\;\;\;\;\;\;\;\;\;(1) 
$$
where $\delta $ is the space quantum. We have used this formula in the
calculation of planetary precession of their perihelions and got quite good
result[7]. Here in this paper, we shall see it also gives satisfactory
explanation to the deflection of the light ray passing at the edge of the
Sun.

The orbital equation in classical physics for centered force is [8]%
$$
h^2u^2(\frac{d^2u}{d\theta ^2}+u)=-\frac Fm\;\;\;\;\;\;\;\;\;\;\;\;(2) 
$$
where $r=\frac 1u$ and $\theta $ are the polar coordinates of the photo, $%
h=cR$, $c$ is the speed of light and $R$ is the radius of the Sun.
Substituting (1) into (2), we get

$$
h^2u^2(\frac{d^2u}{d\theta ^2}+\,u)=\frac{GMu^2}{(1-\delta u)}%
\;\;\;\;\;\;\;\;(4) 
$$
This in the first order turns into

$$
h^2u^2(\frac{d^2u}{d\theta ^2}+\,u)=GMu^2(1+\delta u)\;\;\;\;\;\;\;\;\;(5) 
$$
It follows that

$$
\frac{d^2u}{d\theta ^2}+\,(1-D\delta )u=D\;\;\;\;\;\;\;\;\,\,\,\,\,\,\;(6) 
$$
where $D=\frac{GM}{h^2}$. It is straightforward to verify that the solution
of equation (6) is%
$$
u=A\cos \sqrt{1-D\delta }\theta +B\sin \sqrt{1-D\delta }\theta +\frac
D{1-D\delta }\;\;\;\;\;\;\;\;\;(5) 
$$
in which the constants $A$ and $B$ can be determined in the following
consideration. When $\theta =0$, $r=\infty $, so that $u=0$. This leads to $%
A=-\frac D{1-D\delta }\,$. From $y=r\sin \theta $ and (5) we have%
$$
\frac 1y=\frac D{1-D\delta }\frac{1-\cos \sqrt{1-D\delta }\theta }{\sin
\theta }+B\frac{\sin \sqrt{1-D\delta }\theta }{\sin \theta }%
\;\;\;\;\;\;\;\;\;\;(6) 
$$
when $\theta \rightarrow 0,\;y\rightarrow R$ . It follows that $B=\frac 1{R
\sqrt{1-D\delta }}$ . Thus finally the orbital equation is%
$$
u=\frac D{1-D\delta }(1-\cos \sqrt{1-D\delta }\theta \,)+\frac 1{R\sqrt{%
1-D\delta }}\sin \sqrt{1-D\delta }\theta \;\;\;\;(7) 
$$
Designating the final polar angle of the light ray as $\phi $ , when $%
r\rightarrow \infty $\text{, }the deflection angle may be expressed as $%
\triangle $$\theta =\phi -\pi $ . Thus the equation for $\phi $ is%
$$
\frac D{1-D\delta }(1-\cos \sqrt{1-D\delta }\phi \,)+\frac 1{R\sqrt{%
1-D\delta }}\sin \sqrt{1-D\delta }\phi =0\;\;\;\;\;\;(8) 
$$
This gives%
$$
\phi =\frac 2{\sqrt{1-D\delta }}\left[ \arctan (-\frac{\sqrt{1-D\delta }}{RD}%
)+m\pi \right] \;\;\;\;\;\;\;\;\;(9) 
$$
where $m$ is any integer.

\begin{table}
  \begin{tabular}{|l|l|l|l|l|} \hline
     $\delta $  &  1R  &  1.3R  &  2R & observation \\ \hline
     $\triangle \theta $       &  1.563       &  1.769   & 2.250 & 1.775$\pm $0.019  \\ \hline
  \end{tabular}
  \caption{Deflection of Light Ray near the Sun (unit: second. m=1)\label{key}}
\end{table}

In determining the constant $B$ , we suppose $\theta \rightarrow 0$ . This
is equivalent to presuming $R\rightarrow 0$ . In this arithmetical process,
the space quantum $\delta $ has been specified, somehow inadvertently. This
is in complete accordance with the physical meaning of the space quantum:
the unmeasurable quantity presupposed to be zero in the problem. This has
great exemplary significance in determining the uncertainty quantum which is
crucial in our theoretical framework. In Table 1 we show our calculation for 
$\triangle \theta $ with $\delta =R$ , $1.3R$ and $2R$, together with
experimental observation[9]. It is easily seen that our calculation gives
quite satisfactory explanation to the deflection of the light ray.

Another interesting and perhaps also important discovery in the calculation
is that $m=1$ is only integer giving reasonable $\triangle \theta $ value.
From Table 2 it is easy to find that other values for $m$ produce incredibly
large values which, oddly enough, are symmetrical relative to $m=1$ ,where
there occurs a sudden dramatical fall in the order of magnitude. I believe
this is a profound reflection of its innate quantum nature of the problem,
and therefore, an indication that our framework is reasonable.

\begin{table}
  \begin{tabular}{|l|l|l|l|l|l|l|l|l|l|} \hline
     m  & -7 & -6 & -5 & -4 & -3 & -2 & -1 & 0 & 1  \\ \hline
     $\triangle \theta $   & -10.0a  & -9.0a & -7.8a & -6.5a & -5.2a & -3.9a & -2.6a & -1.3a & 1.769  \\ \hline
     m  & 9  & 8 & 7 & 6  & 5 & 4  & 3  & 2  & 1  \\ \hline
     $\triangle \theta $  & 10.0a & 9.0a & 7.8a & 6.5a & 5.2a & 3.9a & 2.6a & 1.3a & 1.769   \\ \hline
  \end{tabular}
  \caption{Defl. at different values of m. $a=10^6$  (unit: second. $\delta =1.3R$ )\label{key}}
\end{table}

It is one of the most important topics in modern physics to preserve the
quantum feature of quantum mechanics while keeping the theoretical
competence of the theory of relativity. Our research indicate that there may
exist theoretical possibilities other than the theory of relativity.

REFERENCES

1. R. Penrose, {\it The Emperor's New Mind}, (Oxford Univ. Press,1989)

2. S. W. Hawking \& R. Penrose, Sci. American, {\bf 275}, 1, 44(1996)

3. Z. Wang, quant-ph/9605017

4. Z. Wang, quant-ph/9605019

5. Z. Wang, quant-ph/9807035

6. Z. Wang, quant-ph/9806071

7. Z. Wang, quant-ph/9804070

8. Y. B. Zhou, {\it Theoretical Mechanics}, (JiangSu Science Press, 1961)

9. K. R. Lang, {\it Astrophysical Formulae,} (Spriner-Verlag 1974)

\end{document}